\documentclass{svjour3}                     
\smartqed  

\usepackage{graphicx}
\usepackage{mathptmx}      
\usepackage{latexsym}

\newcommand{\msun}{{\rm M_{\odot}}}

\newcommand{\me}{\dot M_{\rm Edd}}

\newcommand{\mo}{\dot M_{\rm out}}

\newcommand{\ngc}{{NGC 4051}}
\newcommand{\pg}{{PG1211+143}}

\newcommand{\einstein}{{\it Einstein Observatory}}

\newcommand{\xmm}{{\it XMM-Newton}}
\newcommand{\chandra}{{\it Chandra}}

\newcommand{\asca}{{\it ASCA}}

\newcommand{\suzaku}{{\it Suzaku}}
\newcommand{\et}{et al.}

\begin{document}

\title{X-ray observations of powerful AGN outflows}
\subtitle{- implications for feedback}

\author{Ken Pounds}

\institute{X-ray and Observational Astrophysics Group\\
           Dept of Physics \& Astronomy\\ 
           University of Leicester \\
           Leicester LE1 7RH, UK \\
              \email{kap@le.ac.uk}}

\date{Received: date / Accepted: date}

\maketitle

\begin{abstract}
Highly ionised winds with velocities $\sim$0.1-0.2c were first detected in X-ray spectra of non-BAL AGN a decade ago. Subsequent observations and
archival searches have shown such winds to be a common feature of luminous AGN, increasing the belief that powerful ionised winds have a wider
importance in galaxy feedback models. Paradoxically, for the best-quantified high velocity outflow (the luminous Seyfert
\pg) the wind appears {\it too powerful} to be compatible with the observed stellar bulge and black hole masses, suggesting the energy coupling of wind to bulge gas
must be inefficient. A recent \xmm\ observation of
the narrow line Seyfert \ngc\ offers an  explanation of this apparent paradox, finding evidence for the fast ionised wind to lose most of
its kinetic energy after shocking against the ISM.  Importantly, the wind momentum is maintained through such a shock, supporting the view that a
momentum-driven flow provides the critical link between black hole and stellar bulge masses implied by the observed M-$\sigma$ relationship.
   
\end{abstract}

\section{Introduction}
\label{intro}
Early X-ray observations of AGN yielded soft X-ray spectra frequently showing the imprint of absorption from ionised gas, the `warm absorber'
(Halpern 1984, Reynolds and Fabian 1995). However, the limited spectral resolution of the \einstein, and \asca\ observations
prevented  important parameters of the warm absorbers, in particular the outflow velocity and mass rate, to be determined with useful precision. The
higher resolution and high throughput afforded by contemporary X-ray observatories, \chandra, \xmm\ and \suzaku\ has transformed that situation
over the past decade, with the warm absorber being shown, typically, to be dominated by K-shell ions of the lighter metals (C, N, O, etc) and
Fe-L, with outflow velocities of several hundred km s$^{-1}$. More dramatic was the detection of blue-shifted X-ray absorption lines in the
iron K band, indicating the presence of highly ionised outflows with velocities $\sim$0.1-0.25c (Chartas \et 2002; Pounds \et 2003; Reeves \et
2003). In addition to adding an important dimension to AGN accretion studies, the mechanical power of  such winds was quickly recognised to have a wider
importance in galaxy feedback models.

Additional detections of high velocity AGN winds were delayed by a combination of the low absorption cross section of highly
ionised gas, and strongly blue-shifted lines in low redshift objects coinciding with falling telescope sensitivity. However, extended observations, particularly with
\xmm, found evidence in 5 additional AGN for outflow velocities of $\sim$0.1-0.2c, with 5 further claims of similar  red-shifted components
(Cappi \et 2006). Residual doubts remained, however, as the majority of detections were of a single absorption line (with consequent uncertainty of
identification), and had moderate statistical significance, raising concerns of `publication bias' (Vaughan and Uttley 2008).  

Those doubts were finally removed  following a blind search of AGN observations in the \xmm\ archive (Tombesi \et 2010), finding 
compelling evidence in 13 (of 42)  radio quiet objects for blue-shifted iron K absorption lines, with implied outflow velocities of
$\sim$0.05-0.25c. A more recent search of the \suzaku\ data archive has yielded a further group of strong detections, with a median outflow
velocity again $\sim$0.1c (Gofford et\ 2013). In addition to confirming that high velocity, highly ionised AGN winds are common, the high yield from these
archival searches shows the flows must typically have a large covering factor, and therefore are likely to involve substantial mass and energy fluxes.

Paradoxically, for the best-quantified high velocity outflow (the luminous Seyfert \pg), in which a wide-angle flow was directly measured (Pounds
and Reeves 2007, 2009), the wind appears {\it too energetic} to be compatible with the observed stellar bulge and black hole masses, suggesting that 
the energy coupling of wind to bulge gas must be inefficient.  

A recent 600 ks \xmm\ observation of the narrow line Seyfert \ngc\  has  provided a possible explanation of that paradox, finding evidence that
the fast ionised wind loses much of its kinetic energy after shocking against the ISM, at a sufficiently small radius for Compton
cooling to be strong (Pounds and King 2013). Crucially, the wind momentum is maintained through such a shock, with repeated events eventually leading to a
momentum-driven thrust as shown (King 2003, 2005) to correctly predict the observed M-$\sigma$ relationship (Ferrarese \& Merritt, 2000; Gebhardt
\et, 2000).

In the following Section, the energetic outflow of \pg\ is reviewed. Section 3 then summarises the essential features of a simple
physical model, based on the \pg\ observation, offering a natural explanation for the existence of highly ionised winds in AGN with
velocities $\sim$0.1c. Section 4 summarises the key results of archival searches showing that such winds are indeed common.
Section 5 reviews the recent observation of \ngc, with evidence for the shocking of such a high velocity wind on collision with 
the ISM. Finally, Section 6 outlines a self-consistent model of the full outflow in \ngc, where the fast wind is shocked at $\sim$0.1 pc, with the
post-shock flow rapidly cooling to yield the characteristic soft X-ray features of the `warm absorber'.

\section{The fast outflow in PG1211+143}

\pg, at a redshift of 0.0809 (Marziani \et\ 1996), is one of the brightest AGN at soft X-ray energies
(Elvis et al 1991). It was classified (Kaspi et al 2000) as a Narrow Line Seyfert 1 galaxy (FWHM H$\beta$ 1800 km~s$^{-1}$), with 
black hole mass  ($\sim4\times10^{7}$M$_{\odot}$) and bolometric luminosity $4\times10^{45}$erg~s$^{-1}$, indicating the mean
accretion rate is Eddington-limited.  

EPIC spectra from an \xmm\ observation of \pg\ in 2001 provided the first evidence for a high velocity ionised outflow in a non-BAL AGN, with
the identication of a blue-shifted Fe Lyman-$\alpha$ absorption line corresponding to an outflow velocity of $\sim$0.09c (Pounds \et 2003). That
observation, closely followed by the detection of a still higher outflow velocity from the luminous QSO PDS 456 (Reeves \et 2003), attracted wide attention
in potentially involving a significant fraction of the bolometric luminosity, and which might be typical of AGN accreting near the
Eddington rate (King and Pounds 2003).

Initial questions were raised about the validity of the high velocity in \pg. The near-coincidence of the observed absorption line blueshift and the redshift of the host
galaxy was a concern, notwithstanding the uncomfortably high column density of heavy metals implied by a local origin. Then, in a detailed modelling of the
soft X-ray RGS data, Kaspi and Behar (2006) found only a much lower velocity. Any doubts relating to the absorption being local were 
removed, however, by a revised velocity of 0.13--0.15c (Pounds and  Page 2006), and when subsequent observations demonstrated the Fe K absorption to be variable (Reeves
\et\ 2008). Furthermore, it now appears that the highest velocity outflows are in general too highly ionised to have significant opacity in
the soft X-ray band (Section 4).

Figure 1 illustrates the initial \xmm\ observation of \pg\ in 2001. The upper panel is a ratio plot of EPIC pn data to a power law continuum,
showing 3 significant `narrow' absorption lines. Identifying the lines at $\sim$7.1 keV, $\sim$2.7 and $\sim$1.5 keV with the resonance
Ly-$\alpha$ transitions of FeXXVI, SXVI and MgXII yields the outflow velocity of 0.09$\pm$0.01c reported in Pounds \et\ 2003. A subsequent
analysis, taking advantage of the higher energy resolution (albeit lower sensitivity) of the EPIC MOS camera, found the absorption features at 
$\sim$1.5 keV and $\sim$2.7 keV to be resolved as line pairs, with the correct energy spacing of the respective He-$\alpha$ and Ly-$\alpha$ resonance
lines of Mg and S.   Additional narrow absorption features match the same K-shell resonance line pairs of Ne, Si and possibly Ar, where - crucially -
all the line energies (figure 1, mid panel) scale from the 7.1 keV line (if now identified with the FeXXV resonance line) with the same observed 
blueshift (with respect to the observer) of $\sim$ 0.05, yielding a revised outflow velocity in the \pg\ rest frame of v$\sim$0.12c (Pounds and
Page 2006).   

The lower panel of figure 1 illustrates absorption in a photoionised gas (derived from the XSTAR code 
of Kallman et al 1996), modelling the MOS spectrum with column density N$_{H}$$\sim$$2\times 10^{22}$ cm$^{-2}$ and ionisation parameter of 
log$\xi$=2.9$\pm$0.4 erg cm s$^{-1}$. The strongest absorption 
lines (in order of increasing energy) correspond to K-shell
resonance transitions of Ne, Mg, Si, S, Ar and Fe. The apparent blueshift read from the XSTAR model of 4.9$\pm$0.3$\times10^{-2}$ was consistent
with  the individual line fitting in corresponding to an outflow velocity (in the AGN rest frame) of v$\sim$0.13$\pm$0.01c.

\begin{figure}                                                                                                
\centering                                                              
\includegraphics[width=6.5cm, angle=270]{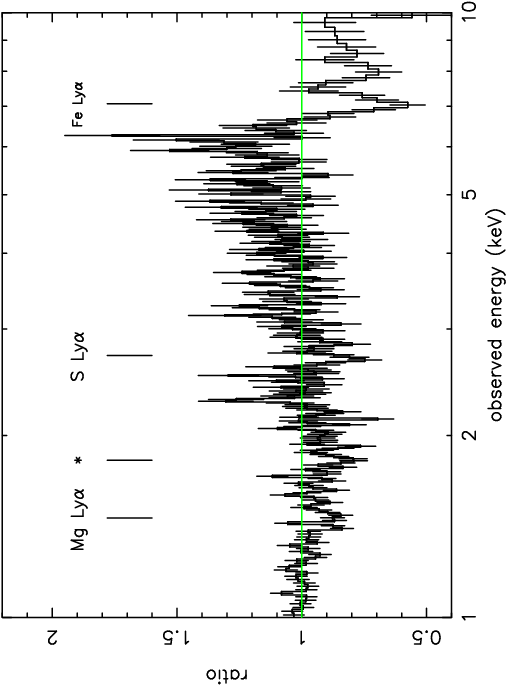}                                                                                  
\centering                                                              
\includegraphics[width=6.5cm, angle=270]{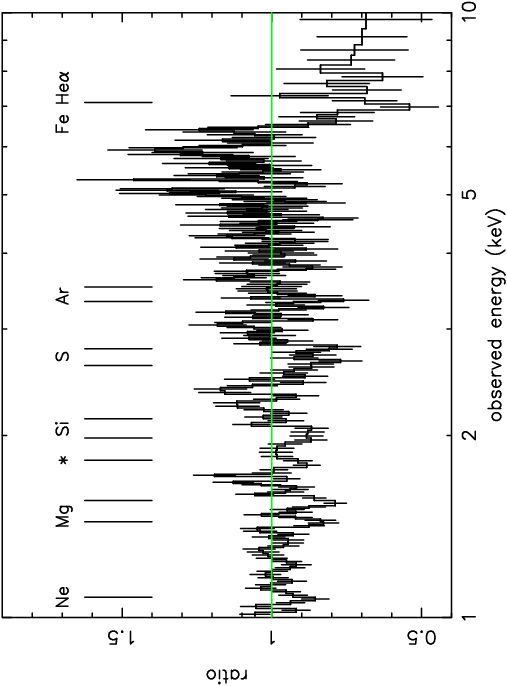}                                                                                  
\centering                                                              
\includegraphics[width=4.7cm, angle=270]{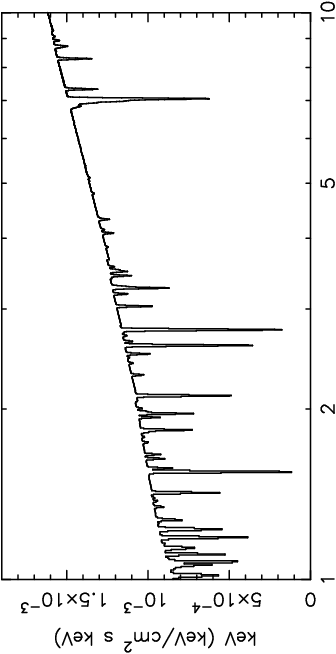}                                                                                  
\caption                                                                
{(top panel) Ratio of EPIC pn spectrum of \pg\ to a power law continuum. Absorption lines are identified with Ly-$\alpha$ of Mg, S and Fe,
with Fe Ly-$\beta$ at $\sim$8 keV. (mid panel) EPIC MOS data resolving H- and He-like ion pairs for a revised blue shift. (lower 
panel) Photoionised absorber model fit to MOS spectrum showing principal K-shell absorption lines of Ne, Mg, Si, S and Ar,
together with K- and L- shell lines of Fe}        
\end{figure} 

\subsection{Mass rate and mechanical energy in the \pg\ outflow}

Although the detection of high speed winds in a substantial fraction of bright AGN (Section 4) suggests most such flows 
have a large covering factor, a
wide angle flow has only been demonstrated directly to date for \pg. 

Using stacked data from 4 \xmm\ observations between 2001 and  2007, Pounds
and Reeves (2007, 2009) examined the relative strength of ionised emission and absorption features to estimate the flow collimation. Analysis of the FeXXV 
PCygni profile (figure 2) indicated a collimation factor b (=
$\Omega$/4$\pi$) of 0.75$\pm$0.25.
The emission line component in the Fe K profile in a 2005 Suzaku observation  appears very similar to that in the stacked \xmm\ data, with a mean
energy of $\sim$6.5 keV and width of $\sigma$$\sim$250eV (Reeves \et\ 2008)).  The
assumption of velocity broadening in a radial flow corresponds to a flow cone of half angle $\sim$50 $\deg$ and b$\sim$0.3.

\begin{figure}                                                                              
\centering                                                              
\includegraphics[width=7cm, angle=270]{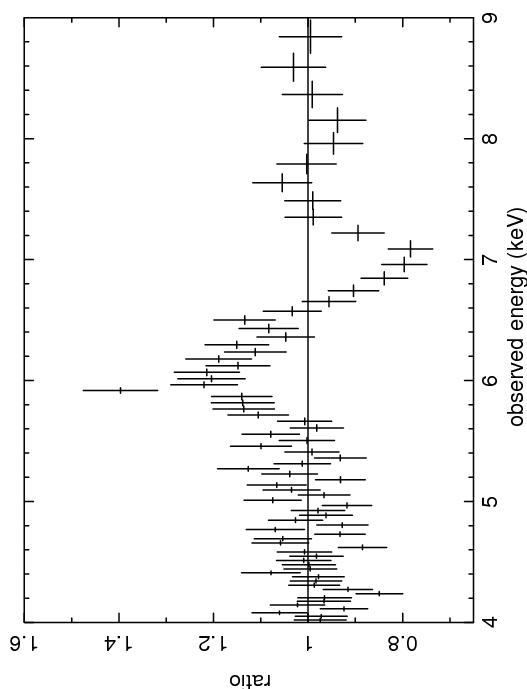} 
\caption                                                                
{The PCygni profile of Fe XXV from the stacked \xmm\ pn observations of \pg\ shows the emission and blue-shifted absorption characteristic
of a wide angle outflow}
\end{figure}

A further estimate, independent of an individual line profile, compared the total continuum energy absorbed and
re-emitted by the highly ionised gas components from modelling the stacked EPIC data (Pounds and Reeves 2009), finding b$\sim$0.5, while a good fit  to the broad
band EPIC spectrum also {\it required} the X-ray emission lines to be broad. 
Taken together, the above indicators confirmed the highly ionised outflow in \pg\ is {\it not} highly collimated. It must therefore involve a significant mass flux, as shown below.

For a uniform radial outflow of velocity v the mass rate is
\begin{equation}
\mo \simeq {4\pi}b{nr^{2}m_p v}, 
\label{massrate}
\end{equation}
where  n is the gas density at a radial
distance r, and m$_{p}$ is the proton mass. 

Modelling the broad band X-ray spectrum of \pg\ provides estimates of the wind ionisation parameter 
($\xi$=L/nr$^{2}$) and the ionising luminosity, yielding nr$^{2}$ $\sim$$3\times10^{39}$ cm$^{-1}$.  For a wind velocity  of 0.12c, and assuming b=0.3, the  mass loss 
rate  
$\mo$$\sim$$7\times10^{25}$ gm s$^{-1}$ ($ \sim 2.5\msun$~yr$^{-1}$), and mechanical energy $\sim$$4.5\times 10^{44}$~erg s$^{-1}$.
 
The mass loss rate is comparable to the Eddington accretion rate 
$\me$ = 1.3$\msun$~yr$^{-1}$ for a supermassive black hole of mass $\sim$$4\times 10{^7}$$\msun$ 
accreting at an efficiency of 10\%. The outflow mechanical energy is $\sim$6$\%$ of the Eddington luminosity, close to that predicted by continuum driving 
(equ. 5 in Section 3 below).   
A simple calculation shows that such a wind is more than sufficient to disrupt star formation in the host galaxy. 

Consider a current episode of Eddington-limited accretion during which
the black hole and bulge masses of \pg\ doubles over $\sim$$3\times 10^{7}$ yr. 
Assuming a mass ratio of M$_{\rm bulge}$ $\sim$ $10^{3}$M$_{\rm BH}$, the binding energy of the bulge gas increases by  E$_{\rm bind}$ $\sim$ M$_{\rm bulge}$$
\sigma^{2}$ $\sim$$2 \times 10^{58}$ ergs (taking $\sigma$=160 km s$^{-1}$ from the M-$\sigma$ relation),
while the high velocity wind injects mechanical energy of $\sim$$4\times
10^{59}$ $\eta$$_{\rm mech}$ erg, where $\eta$$_{\rm mech}$ is the fraction of the wind energy transferred to the bulge gas.
It appears that $\eta$$_{\rm mech}$ $\leq$0.1 for the continued growth of \pg\ not to be disrupted. 

A potential explanation for such low efficiency coupling is presented in Section 5, with new observational evidence for the fast ionised wind in \ngc\ shocking with 
the ISM at a 
sufficiently small radial distance from the AGN for strong Compton cooling to result
in most of the
mechanical energy being lost long before reaching the stellar bulge.

\section{Continuum driving of a highly ionised wind}

The Black Hole Winds (BHW) model (King and Pounds 2003)  provides a simple physical basis for high velocity outflows in AGN
accreting at or above the Eddington limit, and has the bonus of useful predictive power.

The BHW model was inspired by noting that a radial outflow similar to that of \pg\ will be optically 
thick at a sufficiently small radius.
Unit electron scattering optical depth near the base of such a wind will cause each photon to be scattered once on average, before escaping, with the
resulting wind
momentum being of the order of the photon momentum 
\begin{equation}
\mo v \simeq {L_{\rm Edd}\over c},
\label{mom}
\end{equation}
Since
\begin{equation}
\me = {L_{\rm Edd}\over \eta c^2} 
\label{edd}						       
\end{equation}	
where $\eta$ is the accretion efficiency,
the wind velocity 
\begin{equation}
v \simeq {\eta\over \dot m}c \sim 0.1c
\label{v}
\end{equation}
with mechanical energy 
\begin{equation}
1/2\mo v^{2} \simeq {\eta \over 2} {L_{\rm Edd}}
\label{energy}
\end{equation}

The v$^{3}$ dependency of the mechanical energy flux underlines the importance of high velocity for wind feedback, while equation 4 offers a physical basis
on which to compare the observed range of outflow velocities (Section 4).

Assuming the wind is launched and then coasts, the observed outflow velocity will be of order the escape velocity at the launch radius. For \pg\ the wind velocity 
v$\sim$0.12c corresponds to R$_{\rm launch}$ $\sim$ $70R_{\rm s}$  (where $R_{\rm s} = 2GM/c^2$), or $\sim$$8\times10^{14}$~cm.  It is interesting 
to note that EPIC data show significant flux variability in the harder (2-10 keV) band of \pg\ on
timescales of 2-3 hours (fig 1 in Pounds  et al 2003)), compatible with the above scale size relating to the primary (disc/corona) X-ray emission
region.

The results from archival searches presented in Section 4 show that AGN winds with v$\sim$0.1c are indeed common, perhaps indicating that continuum
driving is also common. As only a minority of local AGN appear to be radiating continuously at the Eddington limit (however, see King 2010a),
high velocity winds in such AGN may be intermittent, on timescales perhaps related to the inner accretion disc. The measured
absorption column will be an integral of such a fluctuating wind in the line of sight, while its ionisation state will depend on the current
luminosity.

\section{High speed winds become common}

As noted earlier, the general acceptance of high speed winds in AGN was delayed by concerns relating to the prototype case of \pg.
High velocity outflows in two BAL AGN (Chartas \et\ 2002) and in the most luminous low redshift QSO PDS 456 (Reeves \et\ 2003, O'Brien \et\ 2005)
could be considered rare objects. However, a significant detection of an outflow with velocity $\sim$0.1c was reported for IC4329A (Markowitz \et\
2006), and repeated detections in the range $\sim$0.14--0.2c were found in multiple observations of Mkn 509 (Dadina \et\ 2005). A review in 2006
(Cappi \et 2006) listed 7 non-BAL objects with blue shifts $\sim$0.1c and several with red-shifted absorption lines.   

A major step forward was achieved with the results of the \xmm\ archival search by Tombesi \et\ (2010), finding strong statistical evidence in 15 of 42 radio quiet objects of
blue-shifted iron K absorption lines, with implied outflow velocities up to $\sim$0.3c, and clustering near v$\sim$0.1c (figure 3, top). Subsequent modelling
with XSTAR photoionised grids (Tombesi \et\ 2011) showed the outflows were highly ionised, with log$\xi$ $\sim$ 3--6 erg cm s$^{-1}$,  and had large column densities
in the range N$_{H}$ $\sim$ $10^{22}-10^{24}$ cm$^{-2}$ (figure 3, lower). Among the strongest positive detections was that of \pg, with an 
outflow velocity $\sim$0.15c and an F-test probability of 99.99 $\%$.

\begin{figure*}                                                                                                
\centering                                                              
\includegraphics[width=6.3cm, angle=270]{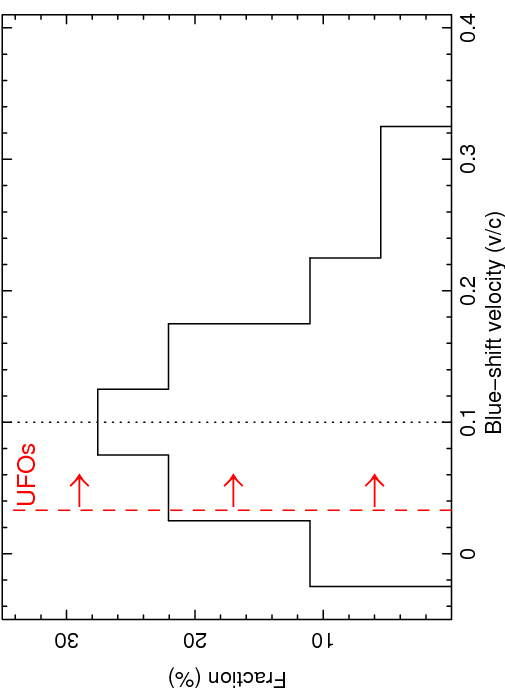}                                                                                                                                                        
\begin{center}
\hbox{
 \hspace{0.2 cm}
   \includegraphics[width=5.6 cm]{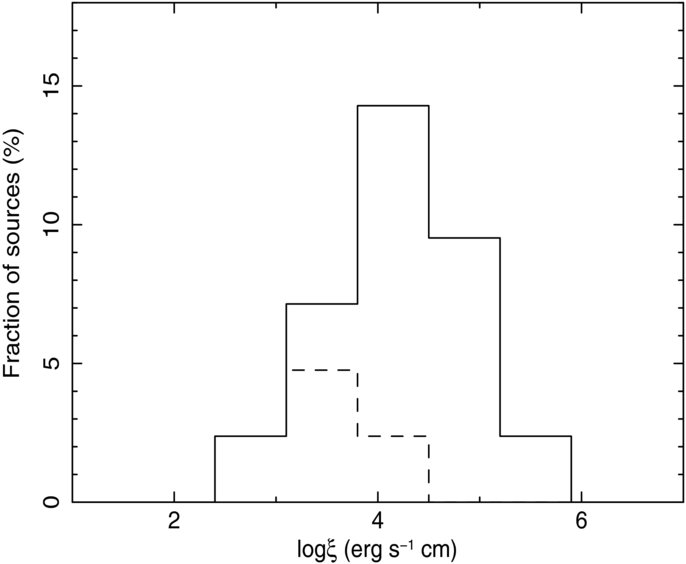}
 \hspace{0.3 cm}
 \includegraphics[width=5.6 cm]{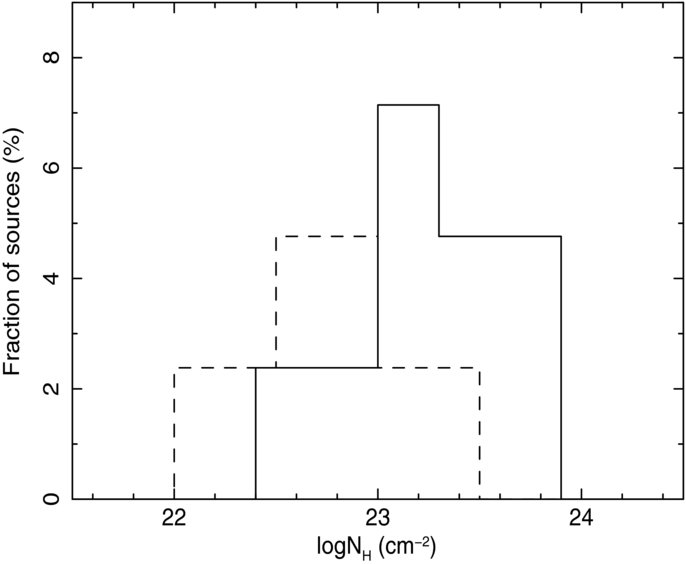}} 
   \end{center} 
\caption
{(top) Distribution of outflow velocities in the Tombesi \et\ sample obtained from examination of extended observations of radio quiet AGN in a
blind search of the \xmm\ data archive. The vertical dotted line indicates the mean value of $\sim$0.1c. (lower) Distribution of ionisation parameter and 
column density from modelling the individual spectra (from Tombesi \et\ 2011). Solid lines refer to the ultra fast outflows, with velocity above $10^{4}$ km s$^{-1}$, and dashed
lines to lower velocity Fe XXV/XXVI absorption} 
\end{figure*}

The high ionisation parameter typical of the fast outflows explains why their detection (in non-BAL AGN) has been restricted to X-ray
observations in the Fe K band, leaving open the additional possibility that fully ionised outflows (quite likely at launch with continuum driving)
will only be detectable when the AGN luminosity falls. As noted above, the observed column density will be a line-of-sight integration over the flow time,
albeit dominated by the higher density at small radii, while the ionisation parameter will be governed by the current AGN luminosity.

A more recent search of the \suzaku\ data archive has yielded a further group of detections, finding significant absorption in the Fe K band in 
20 (of 51) AGN (Gofford \et\ 2013), with velocities up to $\sim$0.3c and a flatter distribution than the \xmm\
sample, but a median value again $\sim$0.1c.  In terms of the
BHW model the higher velocities could relate to a higher value of $\eta$, although it might be premature to suggest such observations as a
reliable measure of black hole spin! Reference to equation 4 also suggests the low velocity tail in both Tombesi \et\ and Gofford \et\
distributions could indicate  a higher accretion ratio, or possibly confusion with a secondary wind that has slowed after being shocked (Section 5).

\section{Evidence for a shocked flow}

The mechanical energy in a fast wind, such as that in \pg, was noted in Section 3 to be incompatible with the continued growth of the black hole and stellar bulge 
in that object, unless the coupling of wind energy to bulge gas is highly inefficient. A recent \xmm\ observation of the narrow line Seyfert
galaxy \ngc\ has provided the first evidence that much of the energy in such a wind may be lost after collision with the ISM.

\ngc\ was found in the \xmm\ archival search to have a high velocity wind, the initial identification with Fe XXVI
Lyman-$\alpha$ in Tombesi \et (2010) indicating a velocity of $\sim$0.15c. Detailed photoionisation modelling subsequently preferred an identification with Fe XXV,
for a velocity v$\sim$0.2c (Tombesi \et\ 2011), the absorption being detected at high significance
(F-test probability 99.95 $\%$) during an observation in 2002 when the source was in an unusually low state. However, in a 2001 observation, 
when the X-ray flux was much higher, no fast wind was seen. 

A 600 ks \xmm\ observation of the Seyfert 1 galaxy \ngc\ in 2009, extending over 6 weeks and 15 spacecraft orbits, revealed a rich absorption spectrum
(figure 4) with outflow velocities,
in both RGS and EPIC spectra, up to $\sim$ 9000 km s$^{-1}$ (Pounds and Vaughan 2011). Strong (and variable) Fe K absorption lines also indicated
outflow velocities in a similar range of $\sim$5000--9000 km s$^{-1}$, while evidence for a higher velocity wind (v$\sim$0.13c) was again
stronger during periods when the ionising continuum was low  (Pounds and Vaughan 2012), suggesting a fast wind that may be fully
ionised at higher continuum levels.  The low redshift (z=0.00234) of \ngc\ also makes a high velocity wind more difficult to detect as the sensitivity of current X-ray
telescopes  is falling sharply above $\sim$7 keV.

\begin{figure}                                                                                                
\centering                                                              
\includegraphics[width=6.5cm, angle=270]{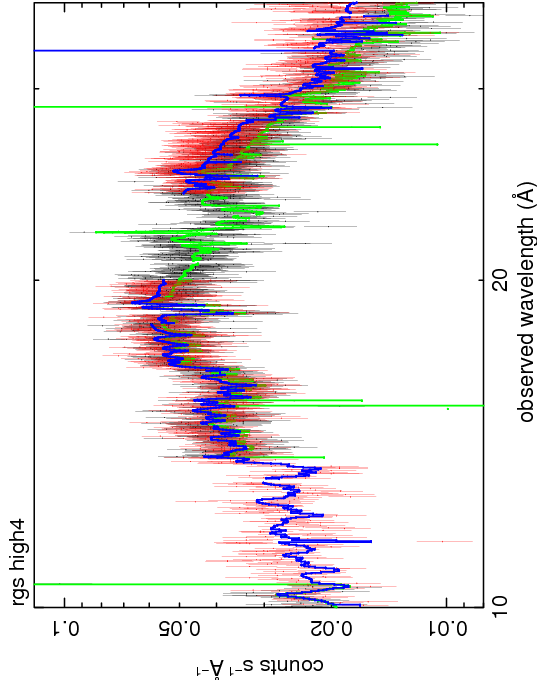}                                                                                  
\centering                                                              
\includegraphics[width=6.5cm, angle=270]{7abs}                                                                                  
\caption                                                                
{XSTAR model fit to the combined RGS1 and RGS2 spectra summed over four successive high-flux orbits from the 2009 \xmm\ observation of \ngc, for the full 10-36 \AA\ 
waveband  in the upper panel, with the section covering strong absorption in OV, VI, VII and VIII highlighted in the lower panel}        
\end{figure}

\begin{figure}                                                                                                
\centering                                                              
\includegraphics[width=7.4cm, angle=270]{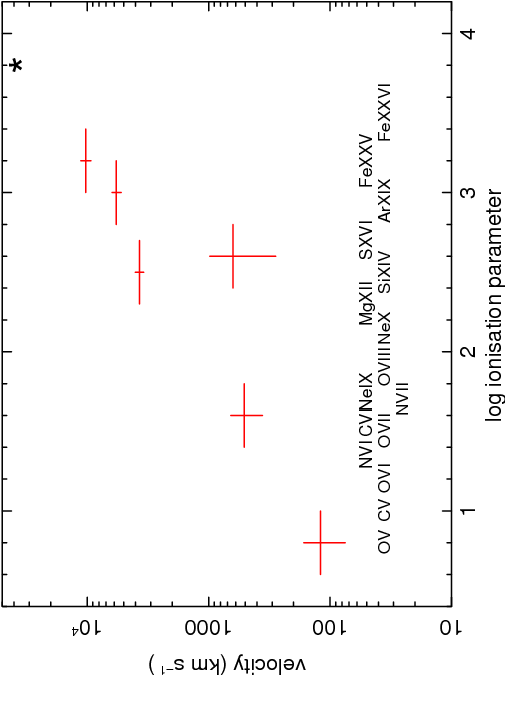}                                                                                  
\caption                                                                
{Outflow velocity and ionisation parameter for each of the XSTAR photoionised absorbers derived from fitting to the RGS and EPIC
spectra,  together with a high point representative of the pre-shock wind, show the linear correlation expected for a
mass conserved  cooling flow (from Pounds and King 2013)}        
\end{figure} 

\begin{figure*}
\begin{center}
\hbox{
 \hspace{0.15 cm}
   \includegraphics[width=4.5cm, angle=270]{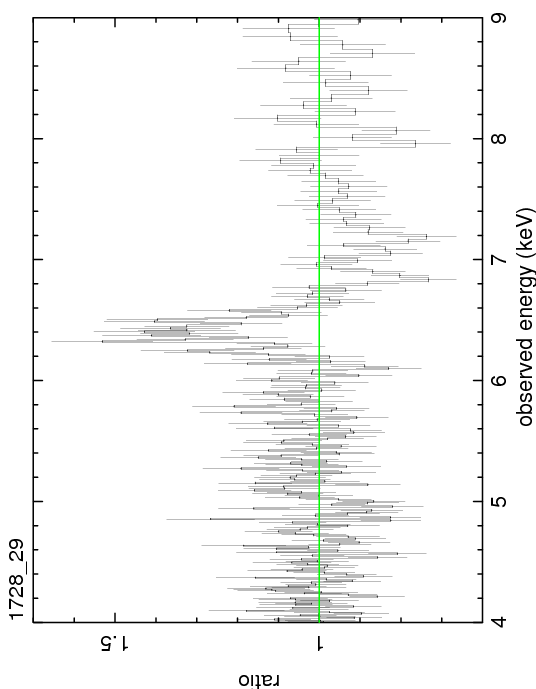}
 \hspace{0.25 cm}
   \includegraphics[width=4.5cm, angle=270]{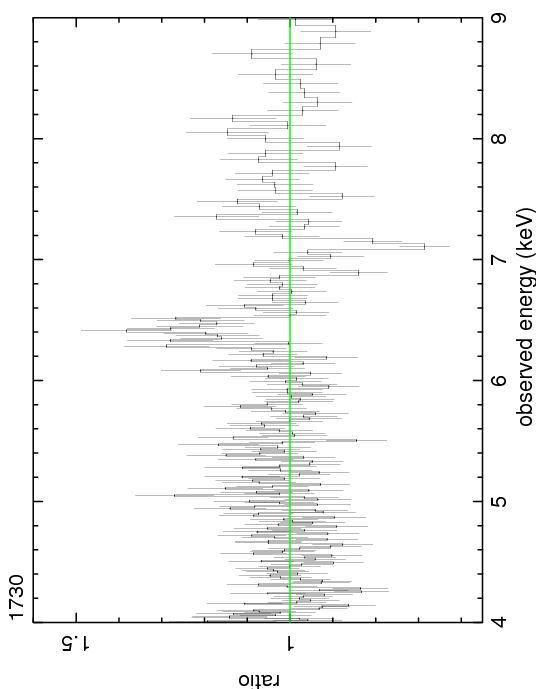}} 
   \end{center} 
\caption
{Fe K profiles for the composite of orbits 6 and 7 and for orbit 8, the latter coinciding with a hard X-ray flare (from Pounds and Vaughan 2012). The ratio of 
resonance absorption lines of Fe XXV and Fe XXVI is a sensitive measure of the ionisation state of the absorbing gas} 
\end{figure*}

\section{A self-consistent model for the shocked wind in \ngc}

More complete modelling of both RGS and EPIC pn absorption spectra of \ngc\ confirmed a general correlation of outflow velocity and
ionisation state (figure 5), as expected by mass conservation in the post-shock flow (King 2010, Pounds and Vaughan 2011). Importantly, the new analysis (Pounds and King
2013) found a large range of column densities to be required by the individual XSTAR 
absorption
components, suggesting an inhomogeneous shocked flow, perhaps with lower ionisation gas clumps or filaments embedded in a more extended,
lower density and more highly ionised flow. 
However, a more important factor in determining the structure of the post-shock flow is likely due to a variable cooling time, as discussed below. 

The key factor in determining the impact of a fast wind is likely to be the distance it travels before colliding with the ISM or previous ejecta.
Compton cooling will dominate for a shock occurring close to the AGN continuum source, with the Compton cooling time and mean flow
speed  then determining the shell thickness of the hotter, more highly ionised  flow component.  However, in the later stages of the flow the  density is likely to increase to the point where two-body cooling becomes important, and then
perhaps dominant.  

At the (adiabatic) shock the free--free and Compton cooling times are
\begin{equation}
t_{\rm ff} \cong 3\times 10^{11}{T^{1/2}\over N}~{\rm s}  = 2{R_{16}^2\over M_8\dot m}~{\rm yr}
\label{ff}
\end{equation}
and
\begin{equation}
t_{\rm C} = 10^{-4}{R_{16}^2\over M_8}~{\rm yr}
\label{compt}
\end{equation}
respectively (see King, Zubovas \& Power, 2011: here $T, N$ are the postshock temperature and number density, $R_{16}$ is the shock radius 
in units of $10^{16}$~cm, $M_8$ is the black hole mass in units of $10^8M_{\odot}$, and $\dot m \sim 1$ is the Eddington factor of the mass 
outflow rate). 

After the adiabatic shock, the gas cools rapidly via inverse Compton cooling, while its density rises as $N \propto T^{-1}$ (isothermal shock 
-- pressure almost constant). So 
\begin{equation}
t_{\rm ff}\propto {T^{1/2}\over N} \propto T^{3/2}, 
\end{equation}
which means that the free--free cooling time decreases sharply while the Compton time does not change. Eventually free--free (and other 
atomic 2--body processes) become faster than Compton when $T$ has decreased sufficiently below the original shock temperature $T_s \sim 1.6\times10^{10}$~K. 
From (\ref{ff}, \ref{compt}) above this requires
\begin{equation}
\left({T\over T_s}\right)^{3/2} < 5\times 10^{-5}
\end{equation}
or 
\begin{equation}
T < 2\times 10^7~{\rm K}.
\end{equation}

The temperature of ionization species forming around a few keV is therefore likely to be determined by atomic cooling processes rather than Compton cooling.
The strong recombination continua in \ngc\ (Pounds and Vaughan 2011a, Pounds and King 2013) are direct evidence for that
additional cooling, with the RRC flux yielding an emission measure  for the related flow component. 

In particular, the onset of strong 2-body cooling will result in the lower ionisation, lower
velocity gas being constrained in a relatively narrow region in the later stages of the post-shock flow. The structure and scale of both high and low ionisation flow
regions can be deduced from the observations and modelling parameters. 

For the highly ionised post-shock flow, the Fe Ly-$\alpha$ to He-$\alpha$ ratio will be governed by the ionising continuum and recombination time. 
Significant variations in this ratio are found on inter-orbit timescales (Pounds and Vaughan 2012), with an example shown in figure 6.
For a mean temperature of $\sim$1 keV, and
recombination  coefficient of $4.6\times10^{-12}$  (Verner and Ferland 1996), the observed recombination timescale of
$\sim$$2\times10^{5}$ s corresponds to an average particle density of $\sim$$4\times10^{6}$ cm$^{-3}$. Comparison with a relevant absorption column
N$_{H}$$\sim$$4\times10^{22}$ cm$^{-2}$ from the XSTAR modelling indicates a column  length scale of $\sim$$10^{16}$ cm. 
Assuming a mean velocity of the 
highly ionised post-shock flow of
6000 km s$^{-1}$, the observed absorption length corresponds to a flow time $\sim$$1.7\times10^{7}$ s (0.6 yr). Equation 7 finds  
a comparable cooling time for \ngc\ at a shock radius R $\sim$$10^{17}$ cm.

For the low ionisation flow component, decay of strong radiative recombination continua (RRC) of NVII, CVI and CV (Pounds and Vaughan 2011a, 
Pounds and King 2013), occurs over $\sim$2-6 days. With an electron temperature from the mean RRC profile of $\sim$5eV, and recombination 
coefficient for CVI of $\sim$$10^{-11}$  (Verner and Ferland
1996), the observed RRC decay timescale corresponds to a (minimum) electron  density of $\sim$$2\times10^{6}$ cm$^{-3}$.  
A column density of 1.5$\times10^{21}$ cm$^{-2}$ from modelling absorption in the main low ionisation flow component then 
corresponds to an absorbing path length of 7$\times10^{14}$ cm.  

Finally, the RRC emission flux provides a consistency check on the above scaling. 
Assuming a solar abundance, and 30 percent of recombinations direct to the ground state, a CVI RRC flux of $\sim$$10^{-5}$ photons
cm$^{-2}$ s$^{-1}$ corresponds to  an emission measure of
$\sim$$2\times10^{62}$cm$^{-3}$, for a Tully-Fisher distance to \ngc\ of 15.2 Mpc. 

\begin{figure*}                                                                                                
\centering                                                              
\includegraphics[width=11cm]{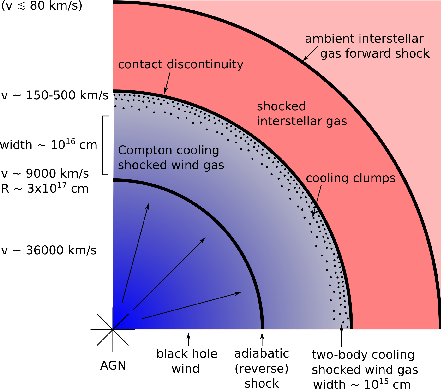}                                                                                  
\caption                                                                
{Structure of the \ngc\ outflow, not to scale, showing the highly ionised wind colliding with the ISM at $\sim$0.1 pc radius, the strong shock
causing a factor 4 drop in velocity. Strong Compton cooling of the shocked gas defines a thin shell where the velocity continues to fall but the ionisation state
remains sufficiently high for strong Fe K absorption. Further along, when 2-body cooling becomes important, the flow rapidly cools and slows over a
narrower region where absorption (and emission) are dominated by the lighter metals seen in the soft X-ray spectrum (from Pounds and King 2013)}        
\end{figure*}

With a mean particle density of $\sim$$2\times10^{6}$cm$^{-3}$ the emission volume (4$\pi.R^{2}$.$\Delta$R) is 
$\sim$$5\times10^{49}$cm$^{3}$ . Assuming a spherical shell geometry of the flow, with fractional solid angle b, shell thickness $\Delta$R
$\sim$$7\times10^{14}$cm, and shell radius R$\sim$$10^{17}$ cm, the measured RRC flux is reproduced for b$\sim$0.5. While this excellent
agreement might be fortuitous given the approximate nature and averaging of several observed and modelled parameters, the mutual consistency 
of absorption and  emission of the photoionised flow is encouraging. Given that only blue-shifted RRC emission is seen, b$\sim$0.5
is consistent with a wide angle flow, visible only on the near side of the accretion disk.

Figure 7 illustrates the main features of the overall \ngc\ outflow, with a fast primary wind being shocked at a radial distance of order 0.1pc, within
the zone of influence of a SMBH of 1.7$\times 10^{6}$ $\msun$. The initially hot gas then cools in the strong radiation field of the AGN, with a
Compton cooling length determining the absorption columns of Fe and the other heavy metal ions. Two-body recombination provides additional
cooling as the density rises downstream, eventually becoming dominant. Absorption (and emission) in the soft X-ray band is
located primarily in this thinner, outer layer of the post-shock flow. 

It is interesting to note that similar shocking of fast outflows provides a natural link beween UFOs (Tombesi
\et\ 2010) and the equally common `warm absorbers' in AGN. While the onset of strong 2-body cooling, resulting in the intermediate
column densities being small, might explain why evidence for intermediate flow velocities has awaited an unusually long observation of a low mass AGN, the accumulated `debris'
of shocked wind and ISM could be a major component of the `warm absorber'.

\section{Summary} 

The requirement of X-ray observations with high sensitivity and high spectral resolution delayed the discovery of powerful,
highly ionised winds from AGN until the launch of \chandra\ and \xmm. A decade after the initial reports, high velocity (v$\sim$0.1c) ionised winds are now established to be common in
low redshift AGN.

The observed distributions of velocity, ionisation parameter and column density are all compatible with winds launched from close to the black
hole, where the optical depth $\tau_{es}$$\sim$1, and carrying the local escape velocity. As the mean luminosity in most low redshift AGN is on average
sub-Eddington, such winds are likely to be intermittent, a view supported by the range of observed column densities.

Recent evidence of a fast wind in \ngc\ being shocked at a distance of $\sim$0.1 pc from the black hole offers an explanation of why such powerful winds
remain compatible with the continued growth of such systems, strong Compton cooling in the AGN radiation field causing most of the wind energy to be
lost before reaching the main star-forming region. Conversely, momentum is conserved through the shock, suggesting a momentum-driven flow is the likely mechanism to 
eventually curtail growth
of the stellar bulge and black hole, as the hole reaches the mass implied by the M-$\sigma$ relation.  

\section{Acknowledgements}
Many thanks to ISSI for the opportunity to take part in a stimulating Workshop; also to Andrew King, Simon Vaughan, Kim Page and James Reeves for helpful
discussions over the period covered by this research.

\end{document}